\documentclass[
  11.5pt,
  abstract=true,
  headings=normal,
  a4paper,
]{scrartcl}
\usepackage[margin=2.0cm
]{geometry}

\usepackage{amssymb}

\usepackage{amsfonts,amsmath,amssymb,amsthm}
\usepackage{bm,nicefrac}
\usepackage{tabularx}
\usepackage{listings}
\usepackage{enumerate}
\usepackage{float}
\usepackage{graphicx}
\usepackage{comment}
\usepackage{subcaption}
\usepackage{paralist}
\usepackage[ruled,lined,algosection]{algorithm2e}
\usepackage[english]{babel}
\usepackage{verbatim}

\usepackage{amsfonts,amsmath,amssymb,amsthm}
\usepackage{mathtools}
\usepackage{graphicx}
\usepackage{hyperref}
\hypersetup{colorlinks,allcolors=blue}
\usepackage{cancel}
\usepackage{siunitx}
\usepackage{url}
\usepackage{paralist}
\usepackage{adjustbox}
\usepackage{orcidlink}
\usepackage{enumitem}
\usepackage{siunitx}

\newcommand{\myref}[2]{\hyperref[#2]{#1~\ref*{#2}}}

\numberwithin{equation}{section}
\usepackage{matlab-prettifier}

\usepackage{tikz}
\newcommand*\circled[1]{\tikz[baseline=(char.base)]{\node[shape=circle,draw,inner sep=2pt] (char) {#1};}}

\def\clap#1{\hbox to 0pt{\hss#1\hss}}

\numberwithin{equation}{section}

\usepackage[affil-it]{authblk}

\usepackage{xpatch}
\xpatchcmd{\author}{\relax#1\relax}{\relax\detokenize{#1}\relax}{}{}
\author{Yi-Kai Kan\thanks{ykan@bnl.gov} }
\author{Derong Xu}

\affil{Brookhaven National Laboratory, Upton, New York, USA}

\addtokomafont{title}{\large}
\begin{document}
\title{Beam-Beam Fields with Full Six-Dimensional Coupling: Theory and Computational Methods}
\date{}

\maketitle

\begin{abstract}
This work presents a generalized theoretical framework for calculating beam–beam fields induced by a Gaussian beam with full six-dimensional coupling. We also develop computational techniques for evaluating these fields. A case study based on the Electron–Ion Collider (EIC) electron storage ring illustrates the practical application of the framework. Our results suggest that the standard slice model for the beam–beam kick can significantly underestimate the longitudinal field. The proposed theory may provide a foundation for developing improved simulation models and guiding future design studies.
\end{abstract}



\section{The Shape Matrix of a Drifting Beam with 6D Coupled Motion in the Head-On Frame}
To model the distribution of a particle beam, we begin with the coordinates of a single particle within the beam. In beam–beam problems, a crossing angle is typically employed between the two colliding beams. To simplify the theoretical analysis of the crossing angle, it is advantageous to transform the problem into the head-on frame using a Lorentz boost. The transformation from the lab-frame coordinates $\mathbf{X}=[\,x,p_x,y,p_y,z,\delta\,]^T$ to the head-on frame coordinates $\mathbf{X}^{*}=[\,x^{*},p^*_x,y^*,p^*_y,z^*,\delta^*\,]^T$ is given by~\cite{hirata1995analysis} 
\begin{equation}
    \begin{cases}
        x^* =  x + z\tan\phi + h^*_z x \sin\phi, \\
        p^*_x = \frac{p_x}{\cos\phi} - h\dfrac{\tan\phi}{\cos\phi},
    \end{cases}
    \begin{cases}
        y^* = y + h^*_y x \sin\phi, \\
        p^*_y = \dfrac{p_y}{\cos\phi},
    \end{cases}
    \text{and }
    \begin{cases}
        z^* = \frac{z}{\cos\phi} + h^*_z x \sin\phi, \\
        \delta^* = \delta -p_x \tan\phi + h\tan^2\phi,
    \end{cases}
\end{equation}
where $\phi$ is the half crossing angle.
If we only keep the linear dynamical variable, the transformation between $\mathbf{X}^{*}$ and $\mathbf{X}$ can be written as
\begin{equation}
    \begin{bmatrix}
       X^*_1\\ X^*_2 \\ X^*_3 \\ X^*_4 \\ X^*_5 \\ X^*_6 
    \end{bmatrix}
    =
    \underbrace{
    \begin{bmatrix}
      1 & 0 & 0 & 0 & \tan\phi & 0 \\
      0 & \frac{1}{\cos\phi} & 0 &0 & 0 & 0\\
      0 & 0 & 1 & 0 & 0 & 0\\
      0 & 0 & 0 & \frac{1}{\cos\phi} & 0 & 0\\
      0 & 0 & 0 & 0 & \frac{1}{\cos\phi} &0 \\
      0 & -\tan\phi & 0 & 0 & 0 & 1
    \end{bmatrix}
    }_{=:\mathcal{L}}
    \begin{bmatrix}
       X_1\\ X_2 \\ X_3 \\ X_4 \\ X_5 \\ X_6 
    \end{bmatrix}.
\end{equation}
Therefore, we can write out transformation for the envelope matrix
\begin{equation}
    \Sigma^{*} = 
    \langle 
    X^{*}(X^{*})^{T}
    \rangle = \mathcal{L}\langle XX^{T} \rangle \mathcal{L}^{T} = \mathcal{L}\Sigma\mathcal{L}^T,
\end{equation}
where $\langle\cdot\rangle$ indicates the mean of a quantity over all particles in the beam. Here, we denote $\Sigma^{*}$ and $\Sigma$ the envelope matrices at $s=0$ in the head-on frame and the lab frame, respectively. In the relativistic regime, the envelope matrix of a drifting particle beam at location $s$ in the head-on frame is given by
\begin{equation}
    \Sigma^{*}(s) :=
    \begin{bmatrix}
        1 & s & 0 & 0 & 0 & 0 \\
        0 & 1 & 0 & 0 & 0 & 0 \\
        0 & 0 & 1 & s & 0 & 0 \\
        0 & 0 & 0 & 1 & 0 & 0 \\
        0 & 0 & 0 & 0 & 1 & 0 \\
        0 & 0 & 0 & 0 & 0 & 1 \\
    \end{bmatrix}
    \begin{bmatrix}
        \sigma^{*}_{11} & \sigma^{*}_{12} & \sigma^{*}_{13} & \sigma^{*}_{14} & \sigma^{*}_{15} & \sigma^{*}_{16} \\
        \sigma^{*}_{21} & \sigma^{*}_{22} & \sigma^{*}_{23} & \sigma^{*}_{24} & \sigma^{*}_{25} & \sigma^{*}_{26} \\
        \sigma^{*}_{31} & \sigma^{*}_{32} & \sigma^{*}_{33} & \sigma^{*}_{34} & \sigma^{*}_{35} & \sigma^{*}_{36} \\
        \sigma^{*}_{41} & \sigma^{*}_{42} & \sigma^{*}_{43} & \sigma^{*}_{44} & \sigma^{*}_{45} & \sigma^{*}_{46} \\
        \sigma^{*}_{51} & \sigma^{*}_{52} & \sigma^{*}_{53} & \sigma^{*}_{54} & \sigma^{*}_{55} & \sigma^{*}_{56} \\
        \sigma^{*}_{61} & \sigma^{*}_{62} & \sigma^{*}_{63} & \sigma^{*}_{64} & \sigma^{*}_{65} & \sigma^{*}_{66} \\
    \end{bmatrix}
    \begin{bmatrix}
        1 & s & 0 & 0 & 0 & 0 \\
        0 & 1 & 0 & 0 & 0 & 0 \\
        0 & 0 & 1 & s & 0 & 0 \\
        0 & 0 & 0 & 1 & 0 & 0 \\
        0 & 0 & 0 & 0 & 1 & 0 \\
        0 & 0 & 0 & 0 & 0 & 1 \\
    \end{bmatrix}^{T}.
\end{equation}
The electromagnetic field generated by a particle beam is mainly characterized by its spatial profile (\emph{i.e.,} its shape). Hence, for the following discussion, we will only focus on the $3\times3$ submatrix of $\Sigma^{*}(s)$:
\begin{equation}\label{eq:shape_matrix}
    \Gamma(s) := \Sigma^{*}(s)\bigl([1,3,5],[1,3,5]\bigr)=
    \begin{bmatrix}
        \sigma^{*}_{11}(s) & \sigma^{*}_{13}(s) & \sigma^{*}_{15}(s) \\
        \sigma^{*}_{13}(s) & \sigma^{*}_{33}(s) & \sigma^{*}_{35}(s) \\
        \sigma^{*}_{15}(s) & \sigma^{*}_{35}(s) & \sigma^{*}_{55}(s)
    \end{bmatrix}.
\end{equation}
The elements of $\Gamma(s)$ are explicitly given  by
\begin{equation}\label{eq:shape_matrix_elems}
\begin{split}
\sigma^{*}_{11}(s) =
&(\sigma_{11} + 2\sigma_{15}\tan\phi + \sigma_{55}\tan^2\phi)
+\frac{2}{\cos\phi}(\sigma_{12} + \sigma_{25}\tan\phi)s
+\frac{\sigma_{22}}{\cos^2\phi}s^2,  \\
\sigma^{*}_{13}(s) = 
&(\sigma_{13} + \sigma_{35}\tan\phi)
+\frac{1}{\cos\phi}(\sigma_{14} + \sigma_{23} +
\sigma_{45}\tan\phi)s
+\frac{\sigma_{24}}{\cos^2\phi}s^2, \\
\sigma^{*}_{15}(s) =
&\frac{1}{\cos\phi}(\sigma_{15} + \sigma_{55}\tan\phi)
+\frac{\sigma_{25}}{\cos^2\phi}s, \\
\sigma^{*}_{33}(s) =
&\sigma_{33}
+\frac{2\sigma_{34}}{\cos\phi}s
+\frac{\sigma_{44}}{\cos^2\phi}s^2, \\
\sigma^{*}_{35}(s) =
&\frac{\sigma_{35}}{\cos\phi}
+\frac{\sigma_{45}}{\cos^2\phi}s, \\
\sigma^{*}_{55}(s) =
& \frac{\sigma_{55}}{\cos^2\phi}. \\
\end{split}
\end{equation}

\section{The Electromagnetic Fields in the Head-On Frame}\label{sec:emfield}
In the previous section, we derive the shape matrix $\Gamma(s)$ in the head-on frame for a Gaussian-distributed strong beam with six-dimensional coupling. The electromagnetic field may be computed directly from this distribution by integrating the corresponding charge density. However, this is generally challenging because the shape matrix $\Gamma(s)$ can be fully coupled, and the result may not be expressed in closed form. Fortunately, an analytical expression for the electric potential exists in the special case of an uncorrelated Gaussian distribution. Therefore, to avoid this complexity, it is advantageous to transform to a coordinate frame in which the distribution becomes decoupled (called decoupled frame). This transformation can be realized by solving an eigenvalue problem for $\Gamma(s)$. Since $\Gamma(s)$ is symmetric, it admits the decomposition
\begin{equation}\label{eq:eigendecomposition_Gamma}
    \Gamma(s) =
    V(s)\Lambda(s)V^{-1}(s),
\end{equation}
where 
\begin{equation*}
    \Lambda(s) =
    \begin{bmatrix}
    \lambda_{11}(s)& 0 &  0&\\
     0& \lambda_{22}(s) & 0&\\
     0 & 0 &\lambda_{33}(s)&\\
    \end{bmatrix}
    \quad\text{and }\quad
    V(s) =
    \begin{bmatrix}
     \mid & \mid & \mid \\
      v_{1}(s)& v_{2}(s) & v_{3}(s)\\
    \mid & \mid & \mid\\
    \end{bmatrix}
\end{equation*}
are the diagonal matrix of eigenvalues and the matrix of eigenvectors, respectively. The quantities $\lambda_{11}$, $\lambda_{22}$, and $\lambda_{33}$ are the squares of the transverse beam sizes and the bunch length in the decoupled frame.

Hence, for the evaluation of electric potential at a position in the head-on frame, we first transform this position $x$, $y$, and $z$ to the decoupled frame
\begin{equation}\label{eq:xform_decoupled}
    \begin{bmatrix}
        \overline{x} \\ \overline{y} \\ \overline{z} 
    \end{bmatrix}
     = 
     \Bigl( V(s) \Bigr)^T 
     \begin{bmatrix}
        x \\ y \\ z
    \end{bmatrix}.
\end{equation}
In earlier treatment of the beam-beam interaction, either the bunch length is assumed to be constant, or the beam distribution in the $x$- and $y$-directions is uncorrelated with $z$-direction. This made it sufficient to use a 2D approximation of the beam-beam potential to calculate the beam-beam fields~\cite{kan2024electromagnetic}. In our formulation, however, both $\overline{z}$ and $\lambda_{33}$ have a dependence on $s$, which may introduce additional terms in the longitudinal field. Therefore, we begin our formulation with the 3D potential. In the decoupled frame, the electric potential generated from a Gaussian beam at a position in the head-on frame can be written as~\cite{kan2024electromagnetic}  
\begin{equation}\label{eq:phi3d}
   \phi^{\text{3D}}\bigl(\overline{x}, \overline{y}, \overline{z}, \lambda_{11}, \lambda_{22},\lambda_{33}\bigr)
    =
    \frac{1}{4\pi\varepsilon_0}\frac{Ne}{(8\pi)^{1/2}}
    \int^{\infty}_{0}
    \frac{
        \exp\left(-\frac{\overline{x}^2}{2(\lambda_{11} + q)}-\frac{\overline{y}^2}{2(\lambda_{22} + q)}-\frac{\overline{z}^2}{2(\lambda_{33} + q/\gamma^2)}\right)
    }{
        (\lambda_{11} + q)^{1/2}(\lambda_{22} + q)^{1/2}(\lambda_{33} + q/\gamma^2)^{1/2}
    }dq,
\end{equation}
where $N$ is the number of particles in the beam, and $e$ is the charge of each particle. The corresponding beam-beam field in the decoupled frame can be computed as
\begin{equation*}
\overline{E}_{x} 
    = -\frac{\partial\phi^{\text{3D}}}{\partial \overline{x}},\quad
\overline{E}_{y} 
    = -\frac{\partial\phi^{\text{3D}}}{\partial \overline{y}},\quad\text{and}\quad
\overline{E}_{z} 
    = -\frac{\partial\phi^{\text{3D}}}{\partial s} 
    -\frac{1}{\gamma^2}\frac{\partial\phi^{\text{3D}}}{\partial \overline{z}},
\end{equation*}
where $\gamma$ denotes the Lorentz factor of the particle beam.
As the variables $\overline{x}$, $\overline{y}$, $\overline{z}$, $\lambda_{11}$, $\lambda_{22}$, and $\lambda_{33}$ all depend on $s$, the longitudinal field $\overline{E}_{z} $ can be further written as
\begin{equation}\label{eq:ez_full_2}
    \overline{E}_z
    =
    \underbracket{ 
        \overbrace{
            -\frac{\lambda^{(1)}_{11}}{2}\frac{\partial^2\phi^{\text{3D}}}{\partial\overline{x}^2} 
            -\frac{\lambda^{(1)}_{22}}{2}\frac{\partial^2\phi^{\text{3D}}}{\partial\overline{y}^2}
        }^{ \circled{1} }
        \overbrace{
            -\overline{x}^{(1)}\frac{\partial\phi^{\text{3D}}}{\partial\overline{x}}
            -\overline{y}^{(1)}\frac{\partial\phi^{\text{3D}}}{\partial\overline{y}}
        }^{ \circled{2} }
    }_{=:\overline{E}^{\text{I}}_z}
    \underbracket{
        \overbrace{
          -\frac{1}{\gamma^2}\frac{\partial\phi^{\text{3D}}}{\partial\overline{z}}
        }^{ \circled{3} }
        \overbrace{
          -\overline{z}^{(1)}\frac{\partial\phi^{\text{3D}}}{\partial\overline{z}}
        }^{ \circled{4} }
    }_{=:\overline{E}^{\text{II}}_z}
    \underbracket{
        \overbrace{
           -\frac{\lambda^{(1)}_{33}}{2}\frac{\partial^2\phi^{\text{3D}}}{\partial\overline{z}^2}
        }^{ \circled{5} }
    }_{=:\overline{E}^{\text{III}}_z}.
\end{equation}
Here, the superscript $^{(k)}$ denotes the $k$-th derivative. The quantities $\overline{x}^{(1)}$, $\overline{y}^{(1)}$ and $\overline{z}^{(1)}$ can be evaluated as
\begin{equation}
    \begin{bmatrix}
        \overline{x}^{(1)} \\ \overline{y}^{(1)} \\ \overline{z}^{(1)}
    \end{bmatrix}=
    \Bigl(V^{(1)}(s)\Bigr)^{T}
    \begin{bmatrix}
        x \\ y \\ z
    \end{bmatrix}.
\end{equation}
Hence, when evaluating Eq.~\eqref{eq:ez_full_2}, we also need to obtain the derivative of the eigendecomposition (Eq.~\eqref{eq:eigendecomposition_Gamma}). Here, we briefly explain each term labeled with a circled number in Eq.~\eqref{eq:ez_full_2}:
\begin{enumerate}[itemsep=0.2em, label=\protect\circled{\arabic*}]
\item This term arises from the hourglass effect and was treated in the absence of coupling in Ref.~\cite{hirata1993symplectic}.  
\item This term originates from the change in the decoupled frame of the test particle's transverse position and was initially addressed in the context of 4D coupling~\cite{leunissen2000six-dimensional}.
\item This term is usually neglected in the beam-beam studies and was systematically investigated in Refs.~\cite{kan2024validation,kan2024electromagnetic}. It was first demonstrated numerically to scale as $\sim 1/\gamma$~\cite{kan2024validation}, and was later shown to be asymptotically bounded above by $ \epsilon^{1/2}/(1 + \epsilon^{1/2})$ through analytical analysis~\cite{kan2024electromagnetic}, where $\epsilon$ is defined as $\epsilon:=\lambda_{22}/(\gamma^2\lambda_{33})$. 
\item This new term reflects the effect of 6D coupling on the test particle’s longitudinal position in the $s$-dependent decoupled frame.
\item This new term emerges as a result of bunch length variations induced by 6D coupling.
\end{enumerate}

In the typical beam-beam problem, the condition $\epsilon:=\lambda_{22}/(\gamma^2\lambda_{33})\to 0$ is usually satisfied. Under such condition, we have the following approximation for the transverse fields
\begin{align}
     \lim_{\epsilon\to 0}\overline{E}_x =&
        -\lim_{\epsilon\to 0}\frac{\partial\phi^{\text{3D}}}  {\partial\overline{x}}
        =-\frac{\partial\phi^{\text{2D}}}  {\partial\overline{x}}
        =:\overline{E}^{\text{2D}}_x, \label{eq:ex}\\
     \lim_{\epsilon\to 0}\overline{E}_y =&
        -\lim_{\epsilon\to 0}\frac{\partial\phi^{\text{3D}}}  {\partial\overline{y}}
        =-\frac{\partial\phi^{\text{2D}}}  {\partial\overline{y}}
        =:\overline{E}^{\text{2D}}_y. \label{eq:ey}
\end{align}
Here, we define a 2D electrical potential
\begin{equation}\label{eq:phi2d}
    \phi^{\mathrm{2D}}(\overline{x},\overline{y},\overline{z},\lambda_{11},\lambda_{22},\lambda_{33}):=\frac{Ne}{8\pi\varepsilon_0}\frac{1}{(2\pi\lambda_{33})^{1/2}}\exp\left(-\frac{\overline{z}^2}{\lambda_{33}}\right)
    \int^{\infty}_{0}
    \frac{\exp\left(-\frac{\overline{x}^2}{2(\lambda_{11}+ q)}-\frac{\overline{y}^2}{2(\lambda_{22} + q)}\right)}{(\lambda_{11} + q)^{1/2}(\lambda_{22} + q)^{1/2}}dq.
\end{equation}
It can be shown that both $\overline{E}^{\text{2D}}_x$ and $\overline{E}^{\text{2D}}_y$ have closed-form expressions~\cite{bassetti1980closed}. 
Therefore, for the longitudinal field, the term $\overline{E}^{\text{I}}_z$ can be readily approximated as
\begin{equation}\label{eq:ezI}
    \lim_{\epsilon\to 0}\overline{E}^{\text{I}}_z =
           \frac{\lambda^{(1)}_{11}}{2}\frac{\partial\overline{E}^{\text{2D}}_x}{\partial\overline{x}} 
          +\frac{\lambda^{(1)}_{22}}{2}\frac{\partial\overline{E}^{\text{2D}}_y}{\partial\overline{y}}
          +\overline{x}^{(1)}\overline{E}^{\text{2D}}_x
          +\overline{y}^{(1)}\overline{E}^{\text{2D}}_y
          .
\end{equation}
On the other hand, the closed-form expression for $\lim_{\epsilon\to 0}\overline{E}^{\text{II}}_z$ and $\lim_{\epsilon\to 0}\overline{E}^{\text{III}}_z$ cannot be easily obtained, since the integrands in $\frac{\partial\phi^{\text{3D}}}{\partial\overline{z}}$ and $\frac{\partial^2\phi^{\text{3D}}}{\partial\overline{z}^2}$ do not satisfy the conditions of the dominated convergence theorem, and hence the limit and the integration cannot be interchanged \cite{kan2024electromagnetic}. In this case, numerical evaluations of the integrals in $\overline{E}^{\text{II}}_z$ and $\overline{E}^{\text{III}}_z$ is necessary, and will be discussed in Section~\ref{sec:ezII}. Once the beam–beam fields in the decoupled frame are computed, the corresponding fields in the original frame can be readily calculated as
\begin{equation}\label{eq:decoupled2headon}
    \begin{bmatrix}
        E_x \\ E_y \\ E_z
    \end{bmatrix}=
    V(s)
    \begin{bmatrix}
        \overline{E}_x \\ \overline{E}_y \\ \overline{E}_z
    \end{bmatrix}.
\end{equation}

\section{Fast Approximation of Shape Matrix Spectral Quantities}\label{sec:fast_approximation}
In the formulation presented in Section~\ref{sec:emfield}, we require the eigenvalues, eigenvectors, and their first derivatives of the shape matrix at each location $s$. The computational cost becomes significant when the beam-beam fields needs to be evaluated at multiple $s$ values, such as in particle tracking. To mitigate this cost, we propose using a second-order Taylor expansion to approximate the eigenvalues and eigenvectors of the $3\times 3$ shape matrix
\begin{equation}\label{eq:decoupled_quantities_taylor}
\begin{split}
    \lambda_{11}(s)&\approx
    \lambda_{11}(0) +
    s\lambda^{(1)}_{11}(0) + 
    \frac{s^2}{2}\lambda^{(2)}_{11}(0), \\
    \lambda_{22}(s)&\approx
    \lambda_{22}(0) +
    s\lambda^{(1)}_{22}(0) + 
    \frac{s^2}{2}\lambda^{(2)}_{22}(0), \\
    \lambda_{33}(s)&\approx
    \lambda_{33}(0) +
    s\lambda^{(1)}_{33}(0) + 
    \frac{s^2}{2}\lambda^{(2)}_{33}(0), \\
    V(s)&\approx
    V(0) + 
    sV^{(1)}(0) +
    \frac{s^2}{2}V^{(2)}(0).
\end{split}
\end{equation}
Once the approximation in Eqs.~\eqref{eq:decoupled_quantities_taylor} is used,  their derivatives can be readily computed as
\begin{equation}\label{eq:decoupled_derivative_quantities_taylor}
\begin{split}
    \lambda^{(1)}_{11}(s)&\approx
    \lambda^{(1)}_{11}(0) + 
    s\lambda^{(2)}_{11}(0), \\
    \lambda^{(1)}_{22}(s)&\approx
    \lambda^{(1)}_{22}(0) + 
    s\lambda^{(2)}_{22}(0), \\
    \lambda^{(1)}_{33}(s)&\approx
    \lambda^{(1)}_{33}(0) + 
    s\lambda^{(2)}_{33}(0), \\
    V^{(1)}(s)&\approx
    V^{(1)}(0) +
    sV^{(2)}(0).
\end{split}
\end{equation}
With this approximation, the eigenvalues, eigenvectors, and their derivatives of the shape matrix $\Gamma(s)$ need to be computed only at $s=0$. This computation can be performed once in advance, and the values at other locations $s$ can then be efficiently obtained using the resulting polynomial expressions.

To illustrate the feasibility of the proposed approximate method, we consider a 4D coupling scenario from EIC electron storage ring. We add a random roll misalignment (in radians) to each quadrupole in the lattice, using the following noise model:
\begin{equation*}
        \varepsilon\sim\mathcal{N}(\mu=0, \sigma=5\times 10^{-4}),\quad \varepsilon\in[-2\sigma,2\sigma].
\end{equation*}
For each random seed, we compute the corresponding coupling matrix $C$ for the lattice as described in~\cite{sagan1999linear}, and then calculate the RMS coupling matrix $C^{\text{RMS}}$ over 3000 seeds. The coupled envelope matrix of the beam at the interaction point (IP) is subsequently calculated from $C^{\text{RMS}}$, and the result is
\begin{equation}\label{eq:coupled_envelope_matrix_ip}
\resizebox{0.8\hsize}{!}{$
    \Sigma=
    \begin{bmatrix}
        8.95 \times 10^{-9} & 0 & -2.11 \times 10^{-10} & 1.46 \times 10^{-9} & 0 & 0 \\
        0 & 4.42 \times 10^{-8} & 3.06 \times 10^{-10} & -1.24 \times 10^{-8} & 0 & 0 \\
        -2.11 \times 10^{-10} & 3.06 \times 10^{-10} & 7.95 \times 10^{-11} & -1.20 \times 10^{-10} & 0 & 0 \\
        1.46 \times 10^{-9} & -1.24 \times 10^{-8} & -1.20 \times 10^{-10} & 2.68 \times 10^{-8} & 0 & 0 \\
        0 & 0 & 0 & 0 & 4.90 \times 10^{-5} & 0 \\
        0 & 0 & 0 & 0 & 0 & 3.02 \times 10^{-7}
    \end{bmatrix}.
$}
\end{equation}
With zero crossing angle ($\phi=0$), the $s$-dependent shape matrix $\Gamma(s)$ (in the head-on frame) can be calculated from Eq.~\eqref{eq:coupled_envelope_matrix_ip} using Eq.~\eqref{eq:shape_matrix} and Eqs.~\eqref{eq:shape_matrix_elems}: 
\begin{equation}
\resizebox{0.9\hsize}{!}{$
    \Gamma(s) = 
    \begin{bmatrix}
        8.95 \times 10^{-9} & -2.11 \times 10^{-10} & 0 \\
        -2.11 \times 10^{-10} & 7.95 \times 10^{-11} & 0 \\
        0 & 0 & 4.90 \times 10^{-5}
    \end{bmatrix}
    +
    s
    \begin{bmatrix}
        0 & 1.77 \times 10^{-9} & 0 \\
        1.77 \times 10^{-9} & -2.40 \times 10^{-10} & 0 \\
        0 & 0 & 0
    \end{bmatrix}
    +
    s^2
    \begin{bmatrix}
        8.84 \times 10^{-8} & -2.48 \times 10^{-8} & 0 \\
        -2.48 \times 10^{-8} & 5.36 \times 10^{-8} & 0 \\
        0 & 0 & 0
    \end{bmatrix}
$}
\end{equation}
\myref{Figure}{fig:eigen_and_derivatives} compares the eigenvalues of $\Gamma(s)$ and their first derivatives obtained from the analytical method described in~\cite{leunissen2000six-dimensional} and from the proposed numerical method (Eqs.~\eqref{eq:decoupled_quantities_taylor} and Eqs.~\eqref{eq:decoupled_derivative_quantities_taylor}). We observe good agreement in the results. However, beyond the 4D coupling, our approach can also handle 6D coupling.

\begin{figure}[H]
\centering
\begin{subfigure}[b]{0.40\linewidth}
    \includegraphics[width=\textwidth]{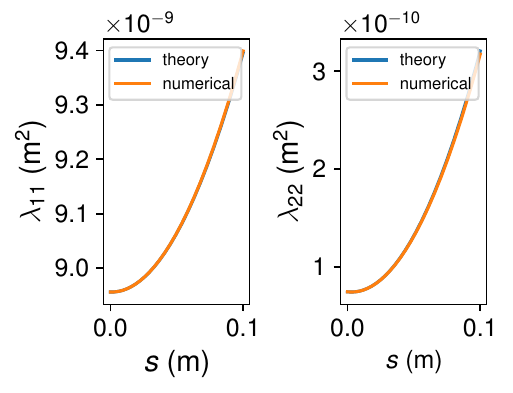}    \caption{\label{fig:eigen_and_derivatives_a}eigenvalues}
\end{subfigure}
\begin{subfigure}[b]{0.40\linewidth}
    \includegraphics[width=\textwidth]{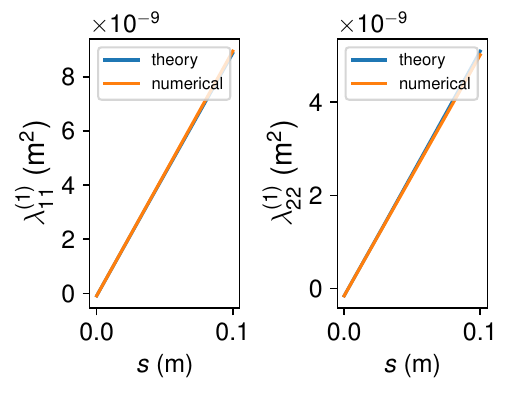}
\caption{\label{fig:eigen_and_derivatives_b}first derivatives of eigenvalues}
\end{subfigure}
\caption{\label{fig:eigen_and_derivatives}Plots of (a) the eigenvalues and (b) their first derivatives for $\Gamma(s)$, computed using the analytical method and the proposed numerical method.}
\end{figure}

\section{Computation of the Derivatives for a Parametric Eigenvalue Problem}\label{sec:deigen}
In the formulation developed in Section~\ref{sec:emfield} or the approximation proposed in Section~\ref{sec:fast_approximation}, it is necessary to compute derivatives of arbitrary order for a parametric eigenvalue problem with respect to a single scalar variable. More specifically, we seek to obtain $\lambda^{(k)}(s)$ and $v^{(k)}(s)$ up to some $k\in\mathbb{N}_{\ge0}$, in which 
\begin{equation}\label{eq:parametric_eigenproblem}
    A(s)v(s) = \lambda(s)v(s).
\end{equation}
This problem has been studied extensively in the literature. In this section, we briefly introduce the formulation summarized in Ref.~\cite{mach2025solving}. We fist consider the Taylor expand for $A(s)$, $v(s)$, and $\lambda(s)$ around some point $s_0$:
\begin{align}
    A(s) =& A^{(0)} + (s-s_0)A^{(1)} +\frac{1}{2}(s-s_0)^2A^{(2)} + \frac{1}{3!}(s-s_0)^3A^{(3)} + \dots, \label{eq:taylor_A}\\
    \lambda(s) =& \lambda^{(0)} + (s-s_0)\lambda^{(1)} +\frac{1}{2}(s-s_0)^2\lambda^{(2)} + \frac{1}{3!}(s-s_0)^3\lambda^{(3)} + \dots, \label{eq:taylor_lambda}\\
    v(s) =& v^{(0)} + (s-s_0)v^{(1)} +\frac{1}{2}(s-s_0)^2v^{(2)} + \frac{1}{3!}(s-s_0)^3v^{(3)} + \dots. \label{eq:taylor_v}
\end{align}
Here, for brevity, we do not write out that all $s$-dependent variables are evaluated at $s_0$. Substituting Eqs.~\eqref{eq:taylor_A}, \eqref{eq:taylor_lambda}, and \eqref{eq:taylor_v} into Eq.~\eqref{eq:parametric_eigenproblem}, imposing the normalization condition $v(s)^Tv(s)=1$, and collecting the coefficients of $(s-s_0)^k$, we obtain~\cite{mach2025solving}
\begin{equation}
\begin{bmatrix}
\begin{array}{c|c}
0 & (v^{(0)})^{T} \\
\hline
v^{(0)} & \lambda^{(0)} I - A^{(0)}
\end{array}
\end{bmatrix}
\begin{bmatrix}
    \lambda^{(k)} \\
    v^{(k)}
\end{bmatrix}
=
\begin{bmatrix}\label{eq:deigen}
    -\frac{1}{2}\sum^{k-1}_{\ell=1}\binom{k}{\ell}\bigr(v^{(k-\ell)}\bigl)^T v^{(\ell)} \\
    \sum^{k-1}_{\ell=0}\binom{k}{\ell}A^{(k-\ell)}v^{\ell}-\sum^{k-1}_{\ell=1}\binom{k}{\ell}v^{(k-\ell)}\lambda^{(\ell)}
\end{bmatrix}.
\end{equation}
Thus, using Eq.~\eqref{eq:deigen}, together with $\lambda^{(0)}$, $v^{(0)}$, and the set $\left\{A^{(\ell)} \,\middle|\, \ell=0,1,\ldots,k\right\}$, we can sequentially compute $\lambda^{(\ell)}$ and $v^{(\ell)}$ for $\ell=1,2,\ldots,k$. Alongside this study, we have developed a package to facilitate these computations~\cite{kan2025deigen}.

\section{Numerical Analysis of the Integrals in the Type II and Type III Longitudinal Fields}\label{sec:ezII}
To evaluate $\overline{E}^{\text{II}}_z$ and $\overline{E}^{\text{III}}_z$,  it is necessary to compute the integrals appearing in $\frac{\partial\phi^{\text{3D}}}{\partial\overline{z}}$ and $\frac{\partial^2\phi^{\text{3D}}}{\partial \overline{z}^2}$, respectively. In this section, we solely focus on the integral in $\frac{\partial\phi^{\text{3D}}}{\partial \overline{z}}$ . Although the integrals in $\frac{\partial^2\phi^{\text{3D}}}{\partial \overline{z}^2}$ differ, the same analysis and technique can still be applied.

Using the change of variable $q:=\lambda_{22}u$, we get the following expression
\begin{equation}\label{eq:integral_in_ez}
    -\frac{\partial\phi^{\text{3D}}}{\partial z}= 
    \frac{1}{4\pi\varepsilon_0}
    \frac{Ne}{(8\pi)^{1/2}}
    \frac{(\lambda_{22})^{1/2}}{(\lambda_{11})^{1/2}(\lambda_{33})^{3/2}}\overline{z}
    \int^{\infty}_{0}
    \underbrace{
    \frac{
        \exp\left(-\frac{\overline{x}^2}{2\lambda_{11}(1 + Au)}-\frac{\overline{y}^2}{2\lambda_{22}(1 + u)}-\frac{\overline{z}^2}{2\lambda_{33}(1 + \epsilon u)}\right)
    }{
        (1 + au)^{1/2}(1 + u)^{1/2}(1 + \epsilon u)^{3/2}
     }
    }_{=:I(u)} 
    du,
\end{equation}
where we define $a:=\lambda_{22}/\lambda_{11}$ and $\epsilon:=\lambda_{22}/(\gamma^2\lambda_{33})$. The integral $I(u)$ in Eq.~\eqref{eq:integral_in_ez} is convergent because it has no singularity at $u=0$ and is asymptotic to $u^{-5/2}$ in the tail region $1/\epsilon \ll u < \infty$. One common approach to handling the numerical integration over the semi-infinite domain is transforming to a finite domain $[0,1]$ by the change of variable $u=t/(1-t)$:
\begin{equation}
\int^{\infty}_{0}I(u)du=
    \int^{1}_{0}
    \frac{
        (1-t)^{1/2}\exp\left(-\frac{\overline{x}^2(1-t)}{2\lambda_{11}(1 - (1-a)t)}-\frac{\overline{y}^2(1-t)}{2\lambda_{22}}-\frac{\overline{z}^2(1-t)}{2\lambda_{33}(1 - (1-\epsilon)t)}\right)
    }{
        (1 - (1-a)t)^{1/2}(1 - (1-\epsilon)t)^{3/2}
     }dt.
\end{equation}
However, since $\epsilon$ is very small, we may encounter catastrophic cancellation in the denominator term $(1-(1-\epsilon)t)^{3/2}$ when $t$ is close to $1$. This can make the result unreliable.

In contrast, the original expression $I(u)$ does not have this issue and is therefore more numerically stable. One way to handle the semi-infinite integration of $I(u)$ directly is choosing a big enough number $M$ such that we numerically integrate over the truncated domain $[0,M]$ using a suitable quadrature method, and the contribution from the tail region becomes negligible. However, there exists an intermediate region for $1\ll u\leq 1/\epsilon$, where $I(u)\sim u^{-1}$ and thus decays slowly. Since $\epsilon$ is typically very small in beam-beam problems, the extent of the intermediate region can be very large. \myref{Figure}{fig:integrand_original} shows the behavior of $I(u)$ using the parameters of the EIC electron storage ring~\cite{willeke2021electron}. From~\myref{Figure}{fig:integrand_original_b}, we observe that the transition from the intermediate region to the tail region occurs very late, around at $u=1/\epsilon\approx 2.58\times 10^{14}$. As a result, accurately integrating over this region may require many quadrature points to achieve reliable results. We alleviate this issue by using a type of double-exponential (DE) transformation~\cite{takahasi1974double,mori2001double} with $u=\exp(\sinh t)$:
\begin{equation}
    \int^{\infty}_{0}I(u)du=
    \int^{\infty}_{-\infty}
    \underbrace{
        I\Bigl(\exp(\sinh t)\Bigr)\cosh t \exp(\sinh t)
    }_{=:\mathfrak{I}(t)}
    dt.
\end{equation}
The resulting tanformed integrand $\mathfrak{I}(t)$ is shown in~\myref{Figure}{fig:integrand_xformed}. The transition from the intermediate region to the tail region now occurs quite early at $t=\operatorname{arcsinh}(\ln (1/\epsilon))\approx 4.19$. This is because the transformation scales the variable at a double-exponential rate. Moreover, we can estimate the asymptotic behavior of $\mathfrak{I}(t)$ in the tail region as
\begin{equation}
    \mathfrak{I}(t)\sim \exp(\sinh t)^{-5/2}\cosh t\exp(\sinh t) 
    \approx \frac{1}{2}\exp\left(-\frac{3}{4}\exp(t) + t\right),
\end{equation}
which indicates the integrand decays double-exponentially. Therefore, it is sufficient to perform the numerical integration over a small truncated interval. The convergence of the integral with respect to the number of trapezoidal points is demonstrated in~\myref{Figure}{fig:de_int_convergence}.

\begin{figure}[H]
\centering
\begin{subfigure}[b]{0.40\linewidth}
    \includegraphics[width=\textwidth]{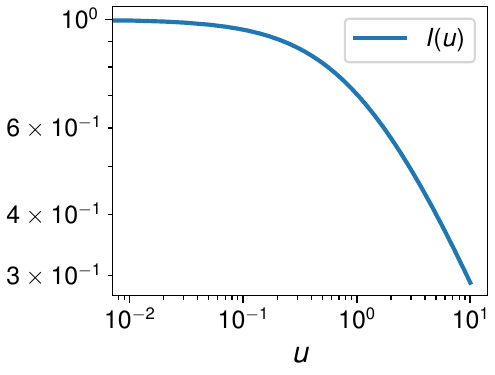}    \caption{\label{fig:integrand_original_a}$I(u)$ over $u\in[0,10]$}
\end{subfigure}
\begin{subfigure}[b]{0.40\linewidth}
    \includegraphics[width=\textwidth]{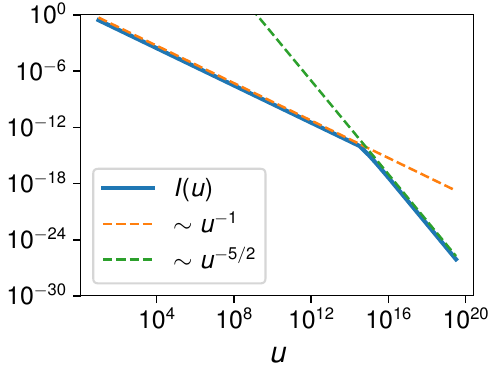}
\caption{\label{fig:integrand_original_b}$I(u)$ over $u\in[10,10^{19}]$}
\end{subfigure}
\caption{Plots of the integrand $I(u)$ over two different domains: (a) $u\in[0,10]$ and (b) $u\in[10,10^{19}]$. The parameters $a\approx 8.09\times 10^{-3}$ and $\epsilon\approx 3.88\times10^{-15}$ used in this calculation are derived from the parameters of the EIC electron storage ring~\cite{willeke2021electron}. \myref{Figure}{fig:integrand_original_b} illustrates the transition from the intermediate region ($\sim u^{-1}$) to the tail region ($\sim u^{-5/2}$), which takes place around $u=1/\epsilon$. }.
\label{fig:integrand_original}
\end{figure}

\begin{figure}[H]
    \centering
    \includegraphics[width=0.4\linewidth]{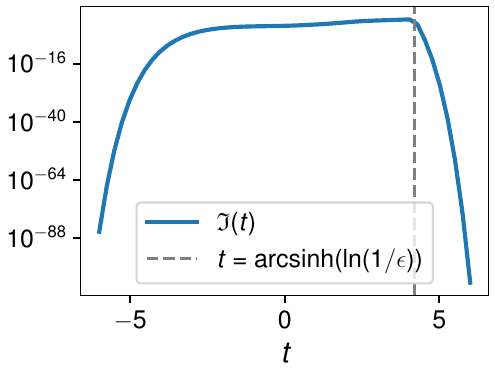}
    \caption{The resulting integrand $\mathfrak{I}(t)$ after the double-exponential transformation. The vertical dashed line marks the transition from the intermediate region to the tail region.}
    \label{fig:integrand_xformed}
\end{figure}

\begin{figure}[H]
    \centering
    \includegraphics[width=0.4\linewidth]{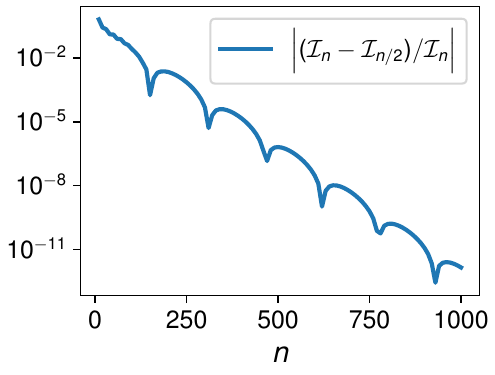}
    \caption{The convergence of the integral with respect to the number of trapezoidal points. Here, $\mathcal{I}_n$} denotes the value of the integral computed using the trapezoidal rule with $n$ quadrature points.\label{fig:de_int_convergence}
\end{figure}

\section{A Case Study}
To demonstrate the results derived from our proposed theory, we consider the case described in Section~\ref{sec:fast_approximation} with a crossing angle of $\SI{25}{\milli\radian}$ (\emph{i.e.,} $\phi=12.5\times10^{-3}$). The corresponding shape matrix is:
\begin{equation}
\resizebox{0.9\hsize}{!}{$
    \Gamma(s)=
    \begin{bmatrix}
        1.66 \times 10^{-8} & -2.11 \times 10^{-10} & 6.13 \times 10^{-7} \\
        -2.11 \times 10^{-10} & 7.95 \times 10^{-11} & 0.00 \\
        6.13 \times 10^{-7} & 0.00 & 4.90 \times 10^{-5}
    \end{bmatrix}
    +
    s
    \begin{bmatrix}
        0 & 1.77 \times 10^{-9} & 0 \\  
        1.77 \times 10^{-9} & -2.40 \times 10^{-10} & 0 \\
        0 & 0 & 0
    \end{bmatrix}
    +
    s^2
    \begin{bmatrix}
        8.84 \times 10^{-8} & -2.48 \times 10^{-8} & 0.00 \\
        -2.48 \times 10^{-8} & 5.36 \times 10^{-8} & 0.00 \\
        0.00 & 0.00 & 0.00
    \end{bmatrix}.
$}
\end{equation}
\myref{Figure}{fig:field_decoupled_a} and \myref{Figure}{fig:field_decoupled_b} show the fields distributions in the decoupled frame at $s=0$ and $s=0.1$, respectively. Both $\overline{E}^{\text{II}}_{z}$ and $\overline{E}^{\text{III}}_{z}$ are negligible in comparison to $\overline{E}^{\text{I}}_z$. Besides, the transverse field $\overline{E}_{x}$ is several order of magnitude bigger than $\overline{E}^{\text{I}}_z$. \myref{Figure}{fig:field_headon_a} and \myref{Figure}{fig:field_headon_b} show the fields distributions in the head-on frame at $s=0$ and $s=0.1$, respectively. The longitudinal field $E_z$ in the head-on frame is 2–3 orders of magnitude larger than in the decoupled frame. Besides, the field pattern of $E_z$ is almost identical to $E_x$. The reason can be seen from the transformation (Eq.~\eqref{eq:decoupled2headon}):
\begin{equation}\label{eq:approximated_decoupled2headon}
    \begin{bmatrix}
        E_x \\
        E_y \\
        E_z
    \end{bmatrix}
    =
    \begin{bmatrix}
     \mid & \mid & \mid \\
      v_{1}(s)& v_{2}(s) & v_{3}(s)\\
    \mid & \mid & \mid\\
    \end{bmatrix}
    \begin{bmatrix}
        \overline{E}_x \\ \overline{E}_y \\ \overline{E}_z
    \end{bmatrix}
    =
    \overline{E}_x v_1(s) + 
    \overline{E}_y v_2(s) +
    (\overline{E}^{\text{I}}_z + \overline{E}^{\text{II}}_z + \overline{E}^{\text{III}}_z) v_3
    \approx \overline{E}_x v_1(s).
\end{equation}
The approximation in Eq.~\eqref{eq:approximated_decoupled2headon} holds because $\overline{E}_y$ vanishes in $x$-$z$ plane, and $\overline{E}_x$ is several orders of magnitude larger than the sum $\overline{E}^{\text{I}}_z + \overline{E}^{\text{II}}_z + \overline{E}^{\text{III}}_z$. This implies that $E_z$ is primarily transmitted from $\overline{E}_x$, and the degree of transmission is determined by the vector $v_1(s)$. In this case, the vectors at $s=0$ and $s=0.1$ are given by:
\begin{equation}
    v_1(s=0)=[0.9996, -0.02378, -0.01250]^T
    \text{ and }
    v_1(s=0.1)=[0.9998, -0.01735, -0.01250]^T.
\end{equation}
It is worth noting that the standard slice model for the beam–beam kick~\cite{hirata1995analysis,leunissen2000six-dimensional} does not include this field projection effect. Our results suggest that this omission could lead to a significant underestimation of the longitudinal field $E_z$.

\section{Conclusion}
In this study, we develop a theoretical framework for calculating beam–beam fields induced by a Gaussian beam with full six-dimensional coupling. We also present computational methods and the necessary numerical analysis for evaluating these fields. Using the EIC electron storage ring parameters as an example, our results suggest that the standard slice model for the beam–beam kick can significantly underestimate the longitudinal field. This limitation arises from the omission of the field projection. The proposed theory also introduces two additional longitudinal fields, $\overline{E}^{\text{II}}$ and $\overline{E}^{\text{III}}$ in the decoupled frame, but their values are negligible compared with $\overline{E}^{\text{I}}$. The proposed framework may provide valuable insights for further investigation.

\begin{figure}[H]
\centering
\begin{subfigure}[b]{\linewidth}
    \includegraphics[width=\textwidth]{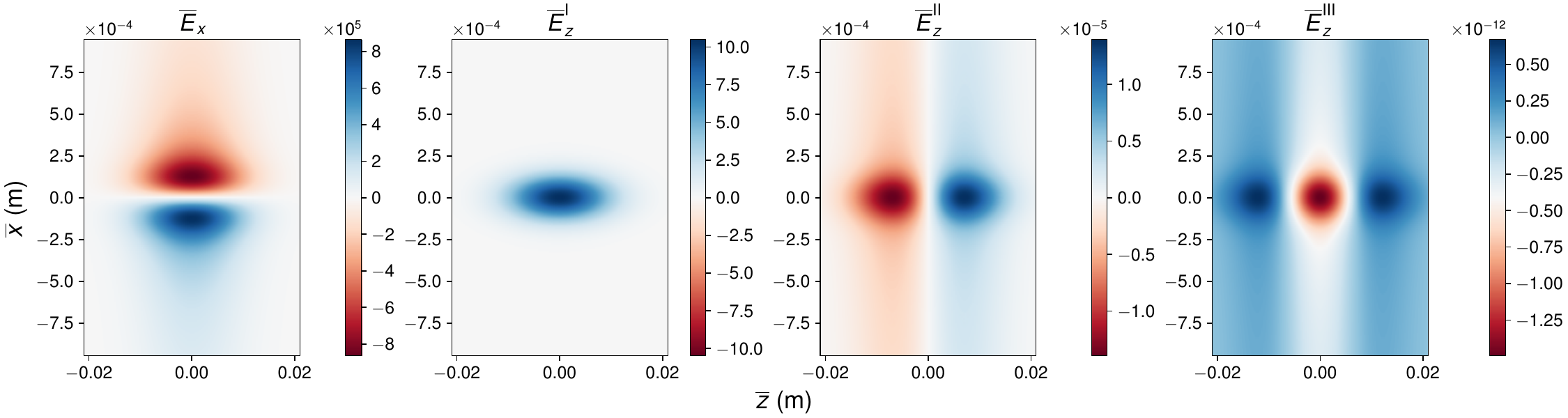}    \caption{\label{fig:field_decoupled_a}field distributions at $s=0$}
\end{subfigure}
\begin{subfigure}[b]{\linewidth}
    \includegraphics[width=\textwidth]{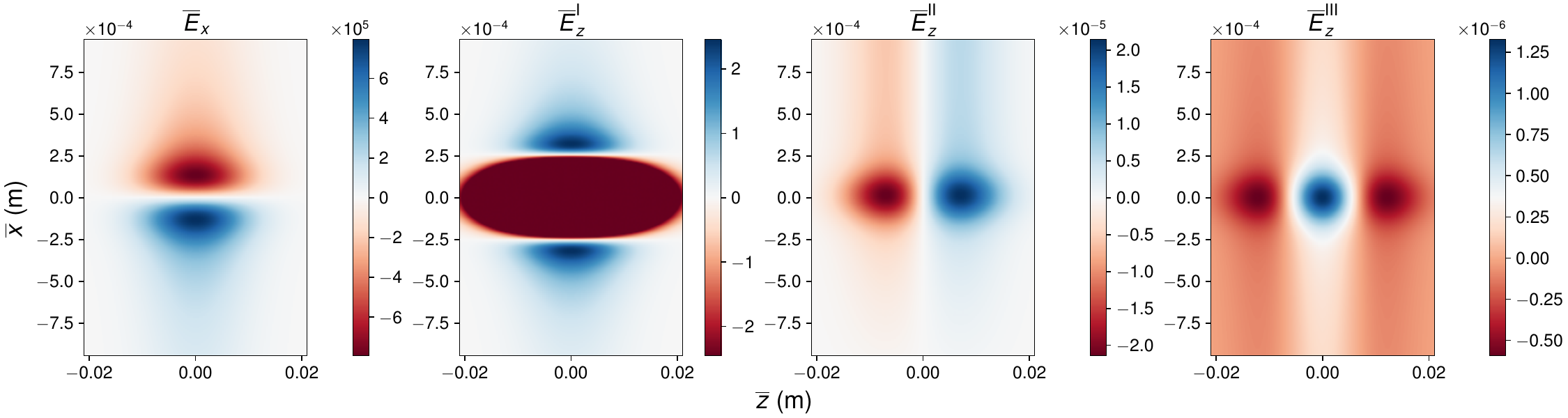}
\caption{\label{fig:field_decoupled_b}field distributions at $s=0.1$}
\end{subfigure}
\caption{Field distribution in the decoupled frame at (a) $s=0$ and (b) $s=0.1$.}.
\label{fig:field_decoupled}
\end{figure}

\begin{figure}[H]
\centering
\begin{subfigure}[b]{\linewidth}
    \centering
    \includegraphics[width=0.60\textwidth]{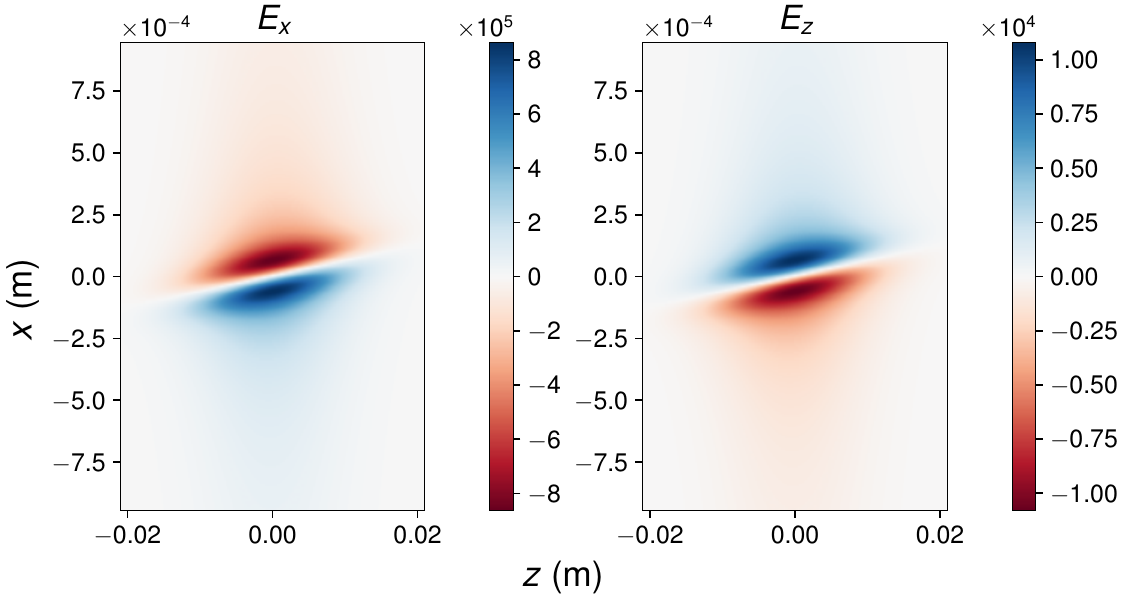}    \caption{\label{fig:field_headon_a}field distributions at $s=0$}
\end{subfigure}
\begin{subfigure}[b]{\linewidth}
    \centering
    \includegraphics[width=0.60\textwidth]{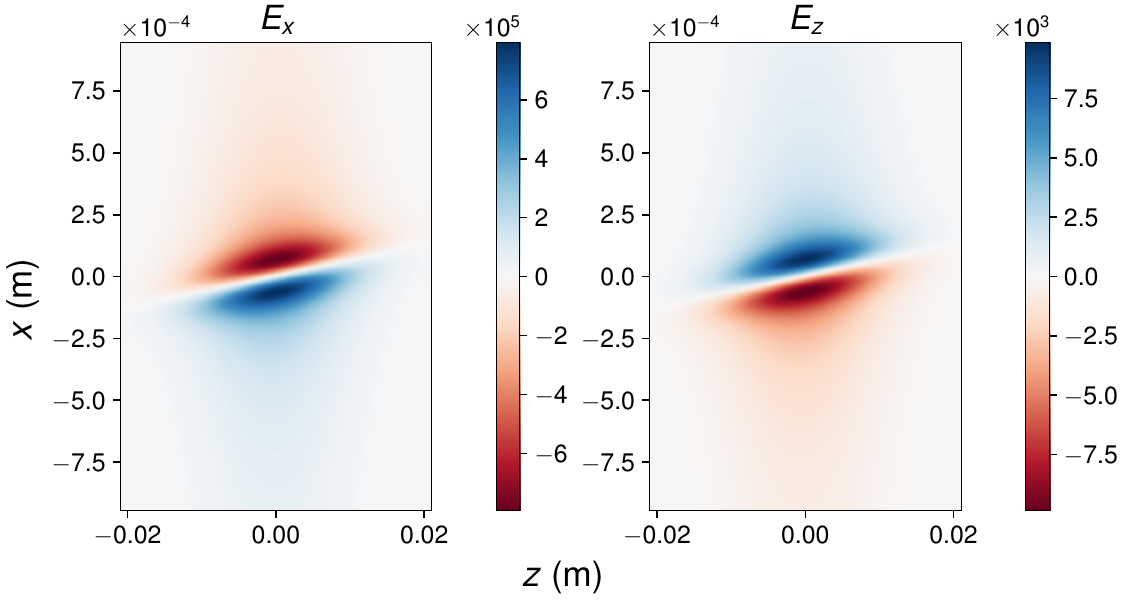}
\caption{\label{fig:field_headon_b}field distributions at $s=0.1$}
\end{subfigure}
\caption{Field distribution in the head-on frame at (a) $s=0$ and (b) $s=0.1$.}.
\label{fig:field_headon}
\end{figure}


\bibliographystyle{elsarticle-num}
\bibliography{references}







\end{document}